\documentclass[
reprint,
superscriptaddress,
floatfix,
amsmath,amssymb,
aps,prl,
]{revtex4-2}

\usepackage{graphicx}
\usepackage{dcolumn}
\usepackage{bm}

\usepackage[pass]{geometry}

\usepackage{amsmath,braket,siunitx}

\usepackage{ulem}
\usepackage{tocvsec2}
\usepackage{graphicx}
\usepackage{hyperref}
\usepackage{booktabs}
\hypersetup{colorlinks=true,breaklinks=true}

\usepackage{xcolor}
\definecolor{myred}{HTML}{e76254}
\definecolor{myblue}{HTML}{0094FB}
\definecolor{mypink}{HTML}{FF73D0}
\definecolor{mygreen}{HTML}{228B22}
\definecolor{darkblue}{HTML}{1e466e}

\AtBeginDocument{%
  \hypersetup{
    citecolor=darkblue,
    linkcolor=darkblue,   
    urlcolor=darkblue}}

\usepackage[usenames,dvipsnames]{xcolor}

\begin{document}

\title{Coexistence of polariton bound states in the continuum and 2D radiative excitons }

\author{Simone Zanotti}
\email{simone.zanotti@ntu.edu.sg}
\affiliation{Division of Physics and Applied Physics, School of Physical and Mathematical Sciences, Nanyang Technological University 637371, Singapore}

\author{Marco Liscidini}
\affiliation{
Dipartimento di Fisica ``A. Volta,'' Universit\`a di Pavia, via Bassi 6, 27100 Pavia (Italy)
}

\author{Dario Gerace}
\affiliation{
Dipartimento di Fisica ``A. Volta,'' Universit\`a di Pavia, via Bassi 6, 27100 Pavia (Italy)
}

\author{Lucio C. Andreani}%
\affiliation{
Dipartimento di Fisica ``A. Volta,'' Universit\`a di Pavia, via Bassi 6, 27100 Pavia (Italy)
}

\begin{abstract}
{Bound states in the continuum (BICs) enable optical modes with ideally infinite radiative lifetimes despite lying within the radiation continuum. Radiation–matter interaction in periodically patterned planar waveguides embedding two-dimensional (2D) optically active excitations can be described quantum mechanically by diagonalizing a non-Hermitian operator, known as the Hopfield matrix, which can be generalized to incorporate independent photonic and excitonic losses into the polaritonic states. However, since 2D excitons undergo intrinsic wavevector-dependent radiative decay within the light cone, whether the Hopfield formalism can consistently account for this process while preserving polariton BICs has remained an open question.
Here we show that a non-Hermitian Hopfield formalism incorporating excitonic radiative losses correctly captures the existence of genuine $k=0$ polariton BICs with diverging radiative lifetimes. 
The theory provides a unified microscopic framework for radiative excitons and polariton BICs, and an efficient predictive tool for designing photonic-crystal platforms coupled to quantum wells, transition-metal dichalcogenides, and other 2D excitonic materials.}
\end{abstract}

\maketitle

\textit{Introduction}.
Strong light-matter coupling between photonic and excitonic degrees of freedom in two-dimensional (2D) materials has emerged as a powerful platform for engineering hybrid quasiparticles, or exciton-polaritons~\cite{Andreani_Book_2014}. In particular, periodically patterned planar waveguides, or photonic crystal (PhC) slabs, embedding 2D electronic excitations of semiconductor materials, such as atomically thin semiconductors or semiconductor quantum wells (QWs), provide a versatile setting for realizing polariton eigenmodes with tailored dispersion, nonlinearities, and topological properties~\cite{Bajoni_PRB_2009,Koshelev_PRB_2018,Zhang_NComm_2018,Chen_NanoLett_2020,HaMyDang_NanoLett_2020,Kim_NanoLett_2021,Ha_M_Dang_AOM_2022}.
Exciton-polaritons can be described on fundamental grounds by a quantum-mechanical theory due to Hopfield~\cite{Hopfield_1958}, which reduces the Hamiltonian problem to diagonalizing a suitable non-Hermitian matrix. However, 2D systems are intrinsically open, {displaying both photonic and excitonic radiative losses,} in addition to possible nonradiative mechanisms. The Hopfield formalism can then be generalized by incorporating photonic and excitonic dissipation {rates} through an imaginary part of their respective energies \cite{Gerace_PRB_2007, Zanotti_PRB_2022}.

An intriguing consequence of radiative interference in open photonic systems is the existence of bound states in the continuum (BICs), i.e., spatially localized modes with energies embedded in the radiation continuum but possessing infinite radiative lifetimes~\cite{Hsu2_NatureRev_2016}. BICs have been extensively investigated in photonic crystals and metasurfaces~\cite{Zhen_PRL_2014,Notomi_PRL_2020}, recently attracting considerable interest in the exciton-polariton context as well~\cite{Koshelev_PRB_2018,Seet2025}, {including anisotropic systems~\cite{art:anisotropic2025},} allowing in particular to achieve long-lived exciton-polaritons displaying either long propagation lengths~\cite{HaMy_Dang_NanoLett_2024}, or inducing unconventional condensed phases of matter~\cite{Ardizzone_Nature_2022,Nigro_PRB_2023, Nigro_PRL_2025,Trypogeorgos_Nature_2025}. 
Whether the non-Hermitian Hopfield formalism can capture the cancellation of radiative decay channels, required for the predictive formation of polariton BICs, has remained unclear so far. In particular, excitonic losses are usually treated as wavevector-independent phenomenological constants~\cite{Gerace_PRB_2007,Zanotti_PRB_2022}, neglecting the momentum-dependent radiative coupling of bare 2D excitons inside the light cone. {Recent theoretical work has investigated the role of coherent interference between excitonic and photonic radiation channels in the formation of polariton BICs~\cite{BoZhen_arxiv_2025}. However, a quantitative description within the generalized Hopfield formalism, incorporating microscopically calculated excitonic radiative losses without phenomenological linewidth parameters, remains to be established.}

In this Letter, we resolve this issue by extending the fundamental quantum Hopfield formalism to include wavevector-dependent radiative losses of 2D excitons in patterned multilayer waveguides. We show that even in the presence of photonic and excitonic decay channels, an exact cancellation of radiation losses leads to $k=0$ polaritonic states with diverging lifetimes, and therefore to genuine BICs for the dressed eigenmodes within the radiative sector. It turns out that polariton BICs and radiative excitons are both present in distinct spectral windows, as they originate from different regions of momentum space. 
{Compared to the experimental observation of polariton BICs \cite{Ha_M_Dang_AOM_2022}, in which the Q-factor was limited by dissipation, here we validate Q-factors in excess of $10^{9}$ in the limit of purely radiative decay.}
Besides providing a microscopic understanding of polariton BIC formation, our approach yields a computationally efficient framework for the design of photonic crystal--exciton architectures, including systems supporting topologically protected polariton states. {Owing to its generality, the formalism is broadly applicable to a description of polariton dispersion including exciton radiative decay in a wide range of realistic material platforms, notably QWs and transition metal dichalcogenides (TMDs).}

\textit{Theory of exciton radiative decay}.
Two-dimensional excitons have intrinsic radiative decay, as first discussed in Refs.~\cite{Agranovich1966,Hanamura1988}. They are radiative when their resonant energy lies within the light cone and form 2D (evanescent) polaritons otherwise. The radiative lifetime of 2D excitons in QWs was first calculated  by neglecting the waveguide effect due to dielectric mismatch~\cite{Andreani1991, Citrin1992, Citrin1993}, and later including it~\cite{Jorda1994}. A non-perturbative treatment of QW polaritons as a function of damping was subsequently given in Ref.~\cite{Creatore2008}. The lifetime was first measured in Ref.~\cite{Deveaud1991}, in agreement with theoretical predictions. The intrinsic radiative lifetime of 2D excitons explains why low-temperature photoluminescence in QWs is dominated by free excitons~\cite{Weisbuch1981}. Similar effects occur in 2D semiconductors, such as TMDs, which have a much larger excitonic oscillator strength, leading to much faster intrinsic radiative decay~\cite{Moody2015, Wang2016, Robert2016}.
Classical calculations of QW reflectivity using local or nonlocal susceptibility treatment were shown to give an excitonic radiative broadening in agreement with quantum theory~\cite{Andreani1991, Ivchenko1992}. A full treatment of radiative and polariton effects in TMDs using nonlocal susceptibility is given in Ref.~\cite{Alpeggiani2018}, focusing on damping and dielectric mismatch. More generally, radiative damping of a Lorentz-Drude dipole arises from interaction with the total electromagnetic field, as recently shown~\cite{Wang-Fan2025}. The interplay between photonic and excitonic radiation channels can furthermore give rise to unusual phenomena, {including recently predicted $k\neq0$ polariton BIC in structures with up-down mirror symmetry~\cite{BoZhen_arxiv_2025}.}

{Recent work has investigated excitonic radiative suppression in van der Waals heterostructure metasurfaces within the weak-coupling regime~\cite{Pantyukhina2025}, and the weak-to-strong coupling crossover using a quasi-normal mode approach~\cite{Fischbach2026}. Despite these developments, a microscopic description of exciton radiative decay integrated into a generalized Hopfield formalism for patterned waveguides remains lacking.} 
{Conventional Hopfield-based approaches typically neglect excitonic radiative losses or introduce them phenomenologically, preventing a fully predictive treatment of photonic and excitonic radiation within this formalism.} To bridge this gap, we hereby develop a quantum formalism yielding the intrinsic radiative decay rate of 2D excitons in patterned waveguides by time-dependent perturbation theory, i.e., by Fermi's Golden Rule. This generalizes the treatment of Refs.~\cite{Andreani1991, Citrin1993, Jorda1994} to arbitrary planar waveguide structures. The derivation is presented in Secs.~S1-S2 of the Supplemental Material (SM). The main outcome is the compact expression
\begin{equation}
\small
\Gamma_\sigma({\bf k}_{\parallel})
=
\Gamma_{0,\sigma}
\left|F({\bf k}_{\parallel})\right|^2
\sum_{j\lambda}
k_jn_j\left|
M_{j,\lambda,\sigma}({\bf k}_{\parallel})
\right|^2\frac{\Theta(k_j-k_{\parallel})
}{\sqrt{k_j^2-k_{\parallel}^2}}
\, ,
\label{eq:gamma-main_compact}
\end{equation}
in which $(j,\lambda)$ identifies a radiative mode outgoing in the lower ($l$) or upper ($u$) cladding with polarization $\lambda=\mathrm{TE,TM}$, $\sigma=x,y,z$ denotes the exciton polarization, ${\bf k}_{\parallel}$ (${k}_{\parallel}=|{\bf k}_{\parallel}|$) is the exciton in-plane wavevector, $k_j=n_j\omega/c$ is the radiation wavevector in cladding $j$, and $\Theta$ is the Heaviside step function selecting the radiative channels. Here, $\Gamma_{0,\sigma}$ is proportional to the exciton oscillator strength per unit area, $F({\bf k}_{\parallel})$ is the excitonic Fourier amplitude, and $M_{j,\lambda,\sigma}$ accounts for the overlap between excitonic and photonic polarizations and mode profiles as defined in the SM.

\begin{figure}[t]
\centering
\includegraphics[width=\columnwidth]{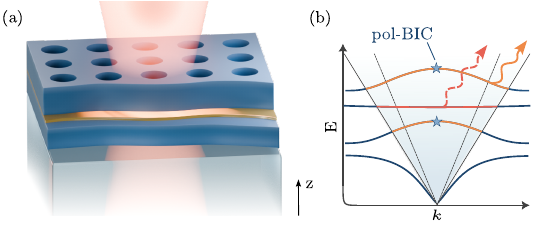}
\caption{Schematic view of the system addressed in this work. (a) A 2D exciton is embedded in a patterned multilayer waveguide (golden layer between the blue ones). It is subject to radiative decay and interacts with the waveguide modes, possibly forming PhC polaritons. (b) Schematic dispersion of the system in (a): Radiative excitons lie above the cladding light line(s) in k-space; PhC polaritons arise from exciton-photon coupling with photonic modes below the light line in the extended-zone scheme and are folded above the cladding light line by periodic patterning.
} \label{fig:structure}
\end{figure}

We further implement the exciton radiative linewidth (\ref{eq:gamma-main_compact}) in the guided-mode expansion (GME) \cite{Andreani_PRB_2006,Minkov_ACSPhot_2020,Zanotti_CPC_2024}, in both its photonic and polaritonic versions, as detailed in Sec.~S2 of the SM. 
This formulation treats photonic and excitonic losses on an equal footing within the Hopfield formalism, constituting the main theoretical advance of this work. It therefore provides a unified description of systems such as that in Fig.~\ref{fig:structure}(a), consisting of an optically active 2D medium embedded in a multilayer waveguide patterned into a photonic crystal. The goal is to capture, within this quantum formalism, the physical processes illustrated in Fig.~\ref{fig:structure}(b): excitons within the light cone of at least one cladding couple to leaky slab modes and become radiative. Exciton states outside both cladding light cones are stationary (neglecting nonradiative decay) and give rise to PhC polariton states folded into the first Brillouin zone by periodic patterning, thus becoming radiative but still forming BICs, as shown below.

\textit{Model system and Methods.}
The physics of the generic system represented in Fig.~1 is hereby analyzed by explicitly considering the multilayered semiconductor structure shown in the End Matter section as a paradigmatic example. The photonic eigenmodes are numerically calculated by the guided-mode expansion (GME) method~\cite{Andreani_PRB_2006, Minkov_ACSPhot_2020}. Polaritonic eigenmodes of the quantum Hamiltonian are derived from a generalized Hopfield matrix diagonalization~\cite{Gerace_PRB_2007,Zanotti_PRB_2022,Zanotti_CPC_2024}, which simultaneously yields the real and imaginary parts as a function of in-plane wave vector. Unlike previous works, here the wavevector-dependent exciton radiative decay is taken into account through Eq.~(\ref{eq:gamma-main_compact}).
The results of the quantum theory are then benchmarked by simulating the same structure with the rigorous coupled-wave analysis (RCWA). Details of both approaches, as well as computational details, are given in {Secs.~S4} and {S5} of the SM. 

\begin{figure}[t]
\centering
\includegraphics[width=\columnwidth]{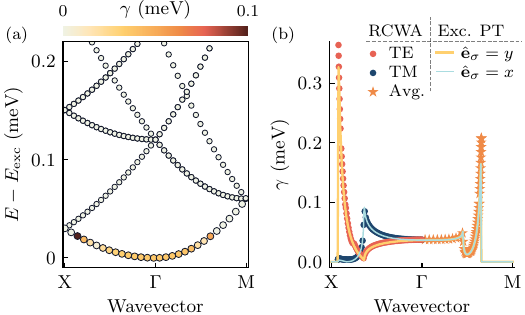}
\caption{(a) Free-exciton dispersion in the first Brillouin zone of a fictitious periodic lattice. The color scale encodes the radiative losses (in meV) calculated using Eq.~\eqref{eq:gamma-main_compact} for excitons polarized along $\hat{\mathbf{e}}_\sigma = y$. Energies are plotted relative to ${E_\mathrm{exc}}=\SI{1.584}{\electronvolt}$. Multiple excitonic bands arise from fictitious band folding; only exciton eigenmodes within the light cone have finite radiative losses ($\Gamma = 2\gamma$), while the folded replicas that are originally outside the light cone retain $\gamma = 0$. (b) Radiative losses of the lowest-energy band in (a), calculated using perturbation theory (PT) and independently extracted from Fano fits of RCWA spectra.
}
\label{fig:exciton}
\end{figure}

\textit{Results: 2D exciton radiative decay.}
The free-exciton dispersion in the effective waveguide is shown in Fig.~\ref{fig:exciton}(a), folded into the first Brillouin zone of the periodic square lattice, as calculated by the extended \texttt{legume} code \cite{Zanotti_CPC_2024}. Excitons above the light line are radiative, as seen by the color scale. Figure~\ref{fig:exciton}(b) shows the radiative linewidth of the lowest excitonic band in (a), calculated by perturbation theory (full lines, see Eq.~\ref{eq:gamma-main_compact}),  
and by fitting RCWA spectra with a Fano lineshape, as explained in Sec.~S4 (points). Excellent agreement is found between the two calculations, thereby validating the perturbative formalism presented in Secs.~S1 and S2. Notice that the radiative linewidth increases upon approaching first the air and then the substrate light line, {owing to the divergence of the} density of states in Fermi's Golden Rule. {Very close to the light line, however}, it drops {again} to very small values, as this divergence is overcompensated by the vanishing electric field factor ({i.e., from the $|h|^2$ factor in} Eq.~S10), yielding a finite radiative linewidth for all wave vectors within the cladding light cones.

\begin{figure}[t]
\centering
\includegraphics[width=\columnwidth]{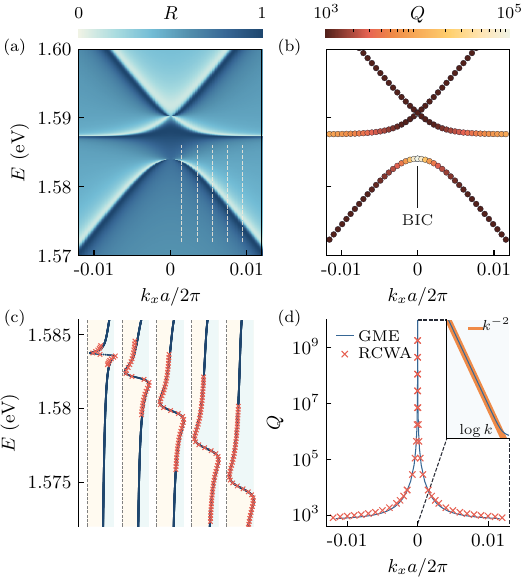}
\caption{(a) Reflectivity spectrum calculated by RCWA for the passive photonic crystal slab
i.e., with the oscillator strength of the active layer set to zero. Vertical dashed lines indicate the spectral cuts in panel (c). (b) Eigenmode dispersion calculated using GME, where the color scale encodes the $k$-dependent $Q$-factor. (c) Spectral cut from panel (a) with the corresponding Fano fit. (d) $Q$-factor of the lowest-energy band calculated using GME and compared with that extracted from RCWA spectra. The divergence at $k=0$ is associated with the symmetry-protected BIC (the expected $1/k_{||}^2$ scaling is shown in the inset, with GME results on a log--log scale).
    }
\label{fig:photonic}
\end{figure}

\textit{Results: photonic eigenmodes.}
The passive photonic crystal structure (i.e., without the excitonic resonance) is calculated by the GME method (in its already mentioned \texttt{legume} implementation), and  benchmarked against RCWA by switching off the light--matter interaction, i.e., by setting the oscillator strength of the active layer to zero ($f/S=0$). Figure~\ref{fig:photonic}(a,b) compares the purely photonic band structure evidenced by clear resonances in the reflectivity spectra calculated by RCWA and the complex eigenmode dispersion obtained by GME, respectively. Both calculations predict a symmetry-protected photonic BIC at normal incidence ($k_{\parallel}=0$) with energy $E_{\mathrm{exc}}=\SI{1.584}{\electronvolt}$. 
{Since we study a supported structure without up-down mirror symmetry, only BICs at $k=0$ are expected \cite{BoZhen_arxiv_2025}.}

The BIC quality factor ($Q$-factor) can be investigated in complementary ways within the two methods. In RCWA, the resonant frequencies and linewidths are extracted from Fano fits of RCWA spectra (see Sec.~S4 of the SM). Examples of these fits are shown in Fig.~\ref{fig:photonic}(c), where each spectrum corresponds to a vertical dashed line in Fig.~\ref{fig:photonic}(a). Approaching the BIC, the scattering signal progressively weakens and eventually vanishes at the symmetry-protected point $k_{\parallel}=0$. Thus, while RCWA can approach the BIC arbitrarily closely by increasing the spectral resolution, the exact BIC point is more naturally captured by GME. In fact, the radiative linewidth is obtained directly from first-order perturbation theory in GME, where the resonance linewidth corresponds to twice the imaginary part of the complex eigenfrequency. As shown in Fig.~\ref{fig:photonic}(d), the $Q$-factors from RCWA and GME are in excellent quantitative agreement, both reproducing the characteristic $Q\propto1/k_{\parallel}^2$ scaling \cite{Koshelev2018_PRL}, as highlighted by the reference line in the inset.

\begin{figure}[t]
\centering
\includegraphics[width=\columnwidth]{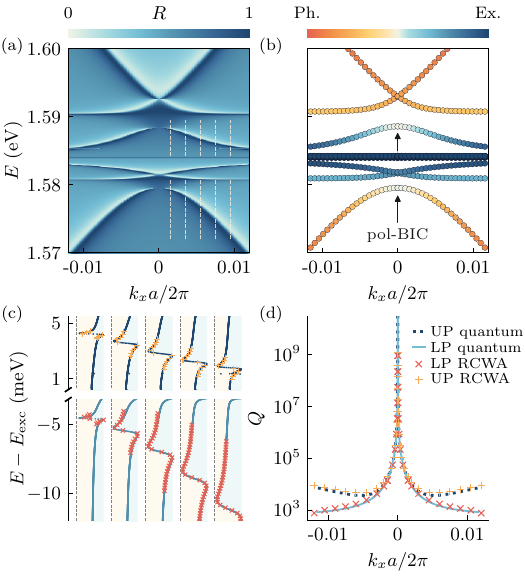}
    \caption{(a) Reflectivity spectrum calculated by RCWA for the same structure as in Fig.~\ref{fig:photonic}, now including the excitonic resonance {with zero nonradiative broadening} ($\gamma_{\mathrm{nonrad}}=0$). Vertical dashed lines indicate the spectral cuts in panel (c). (b) Polariton dispersion calculated using quantum theory (Hopfield matrix diagonalization), with the color scale encoding the excitonic fraction. (c) Spectral cuts from panel (a), together with the corresponding Fano fits. The lower and upper panels show the fitting regions around the lower- and upper-polariton branches, respectively. (d) $Q$-factors of the lower- and upper-polariton branches calculated by quantum theory and compared with those extracted from RCWA spectra, both diverging at $k_{\parallel}=0$ (polariton BICs), despite the presence of intrinsically radiative excitonic states at the same wavevector.
    }
\label{fig:polariton}
\end{figure}

\textit{Results: polariton BICs}.
Finally, we include the excitonic oscillator strength of the active layer in our simulations, allowing the photonic modes to strongly couple with the excitonic resonances . The resulting PhC polariton dispersions, calculated by  RCWA and the generalized Hopfield formalism, are shown in Figs.~\ref{fig:polariton}(a,b), respectively. The color scale in panel (b) encodes the excitonic fraction of each polaritonic branch as directly calculated from the diagonalized Hopfield modes, confirming the hybrid light--matter nature of both the lower- and upper-branches. Excellent agreement is found between the two calculations throughout the dispersion.

As for the purely photonic case, the resonant frequencies and linewidths extracted from RCWA are obtained by fitting reflectivity spectra with Fano lineshapes. Representative fits for the lower- and upper-polariton branches are shown in Fig.~\ref{fig:polariton}(c), while the corresponding $Q$-factors are compared with the generalized Hopfield results in Fig.~\ref{fig:polariton}(d). Remarkably, the agreement remains quantitatively excellent for both branches over the entire wavevector range. However, the existence of a polariton BIC in the diagonalized Hopfield matrix results is far from obvious. In the considered energy range, the photonic mode couples simultaneously to a manifold of excitonic states, including excitons above the light cone, which possess intrinsic radiative decay. These radiative excitons remain visible around the bare exciton energy of $\SI{1.584}{\electronvolt}$, demonstrating that the system simultaneously supports bright excitonic resonances \textit{and} a radiatively decoupled polariton state at $k_{\parallel}=0$.
In fact, the microscopic origin of this behavior is surprisingly well captured by the present theory. The photonic BIC arises from a symmetry-protected superposition of photonic modes with reciprocal lattice vectors $2\pi(\pm1,0)/a$ and $2\pi(0, \pm1)/a$. Through light--matter coupling, this state hybridizes with the excitonic superposition having the same symmetry. Crucially, this excitonic combination is composed of states below the light cone and therefore does not carry intrinsic radiative losses. Consequently, both photonic and excitonic radiative decay channels are simultaneously suppressed, allowing the formation of a true polariton BIC despite the presence of intrinsically radiative excitonic states.
{While the formation of polariton BICs at $k=0$ is consistent with symmetry arguments, the present formalism allows calculating the polariton dispersion and exciton radiative decay at all wavevectors, within a first-principles theory without any fit or adjustable parameters.}
With finite nonradiative excitonic broadening ($\gamma_{\mathrm{nonrad}}$), the polariton BIC also acquires a broadening: detailed results are shown in {Sec.~S6} the SM.

\textit{Final remarks and conclusions.}
In conclusion, we have answered a fundamental question concerning the quantum description of open exciton-photon systems: can a polariton bound state in the continuum exist when two-dimensional excitons from which it originates possess intrinsic radiative decay? By extending the Hopfield formalism to include microscopic, wavevector-dependent excitonic radiative losses in two-dimensional layered structures, we demonstrate that the answer is affirmative. Polariton BICs remain exact eigenstates of the coupled system because their excitonic component consists of non-radiative momentum states outside the light cone, even though bright excitonic resonances coexist at the same energy. This establishes a unified microscopic framework for describing radiative excitons and polariton BICs, providing a predictive foundation for designing high-Q exciton–photonic-crystal platforms, and for future studies of topological and many-body polaritonic phenomena.

\textit{Acknowledgments}. 
Useful scientific discussions with M. Minkov, L. He, and B. Zhen are gratefully acknowledged. S.Z. acknowledges support from the National Research Foundation grant N-GAP (NRF2023-ITC004-001).

\bibliography{biblio}

\clearpage
\newpage

\section*{End Matter}

\textit{Generalized Hopfield matrix}---The radiative linewidth reported in Eq.~\eqref{eq:gamma-main_compact} (and explicitly derived in SM) is  incorporated into the numerical procedure already described in previous works~\cite{Zanotti_PRB_2022,Zanotti_CPC_2024}, which represents a fully quantum theory based on a  generalization of the Hopfield formalism. Following the same derivation, the coupled exciton--photon system at a fixed wavevector $\mathbf{k}_{||}$ is recast as the generalized Hopfield eigenvalue problem, requiring the diagonalization of the following ``Hopfield matrix’’ {, which is distinct from the Hamiltonian and is generally non-Hermitian even in the absence of radiative losses:}
\begin{equation}\label{eq:hopfield}
M=
\begin{pmatrix}
\boldsymbol{\omega}+2\Re\mathbf{D} & -i\mathbf{\Gamma} & -2\Re\mathbf{D} & -i\mathbf{\Gamma}\\
i\mathbf{\Gamma}^\dagger & \mathbf{\Xi} & -i\mathbf{\Gamma}^\dagger &\mathbf{\tilde{0}}\\
2\Re\mathbf{D} & -i\mathbf{\Gamma} & -\boldsymbol{\omega}-2\Re\mathbf{D} & -i\mathbf{\Gamma}\\
-i\mathbf{\Gamma}^\dagger & \mathbf{\tilde{0}} & i\mathbf{\Gamma}^\dagger & -\mathbf{\Xi}
\end{pmatrix} \, .
\end{equation}
The $N\times N$ purely photonic block is given by
\begin{equation}
\boldsymbol{\omega}+2\Re\mathbf{D}
=
[\![ \hbar\tilde{\omega}_n \delta_{n,n'}
+\sum_{\nu,\sigma,\alpha}
2\Re\!\left[D_{nn'\nu\sigma j}\right]
]\!] \, ,
\end{equation}
in which $\hbar\tilde{\omega}_n$ are the complex energies of the photonic modes, and $N$ denotes the number of modes retained in the basis. The index $\alpha\in[1,K]$ labels the active layers, while $\nu\in[1,M]$ and $\sigma\in[x,y,z]$ label the excitonic eigenmodes and their corresponding polarizations. The symbol $\Re$ stands for the real part. The matrix $\mathbf{D}$ accounts for the photonic self-interaction originating from the $\mathbf{A}^2$ term of the light--matter interaction Hamiltonian; its derivation is unchanged from Ref.~\cite{Zanotti_CPC_2024}, and it is therefore not repeated here. Although the contribution of $\mathbf{D}$ is typically small (and often neglected), it is retained in our formalism for completeness.
The coupling and the excitonic blocks are defined as
\begin{equation}
\mathbf{\Gamma}=
\begin{bmatrix}
\mathbf{C}_1 & \mathbf{C}_2 & \cdots & \mathbf{C}_K
\end{bmatrix},
\qquad
\mathbf{\Xi}=
\begin{bmatrix}
\mathbf{E}_1 & \mathbf{0} & \cdots & \mathbf{0}\\
\mathbf{0} & \mathbf{E}_2 & \cdots & \mathbf{0}\\
\vdots & \vdots & \ddots & \vdots\\
\mathbf{0} & \mathbf{0} & \cdots & \mathbf{E}_K
\end{bmatrix} \, ,
\end{equation}
which are in turn composed of the matrices
\begin{equation}
\mathbf{C}_\alpha=[\![C_{n,\nu,\sigma}^\alpha]\!],
\qquad
\mathbf{E}_\alpha=
[\![\tilde{E}_{\nu\sigma}^\alpha
\delta_{(\nu\sigma),(\nu'\sigma')}]\!] \, .
\end{equation}
The coupling coefficients $C_{n,\nu,\sigma}^\alpha$ are calculated by following the formalism described in Ref.~\cite{Zanotti_CPC_2024}. In that work, the complex excitonic energies were taken as
\begin{equation}
\tilde{E}_{\nu\sigma}^\alpha
=
E_{\nu\sigma}^\alpha
+i\gamma_{\mathrm{nonrad}} \, ,
\end{equation}
in which only a dispersionless non-radiative loss $\gamma_{\mathrm{nonrad}}$ was included. The formalism developed in the present work generalizes the latter description by incorporating the wavevector-dependent radiative linewidth expressed in Eq.~\eqref{eq:gamma-main_compact}, such that
\begin{equation}\label{eq:ex_loss}
\tilde{E}_{\nu\sigma}^\alpha
=
E_{\nu\sigma}^\alpha
+i\gamma_{\nu\sigma}^\alpha
+i\gamma_{\mathrm{nonrad}} \, ,
\end{equation}
in which {$\gamma_{\nu\sigma}^\alpha=\Gamma_{\nu\sigma}^\alpha/2$} is derived from Eq.~\eqref{eq:gamma-main_compact}. The diagonalization of the full Hopfield matrix in Eq.~\eqref{eq:hopfield} above, with the modified excitonic block $\mathbf{\Xi}$, provides the complete description of radiative exciton--polaritons, and correctly accounts for the polariton bound states in the continuum, as shown in the results reported in the main text. {Complementing recent work employing an effective model fitted near the $\Gamma$ point~\cite{Pantyukhina2026}, the Hopfield matrix in Eq.~\eqref{eq:hopfield} incorporates first-principles excitonic radiative decay rates calculated at arbitrary in-plane wavevectors.}

\begin{figure}[t]
\centering
\includegraphics[width=\columnwidth]{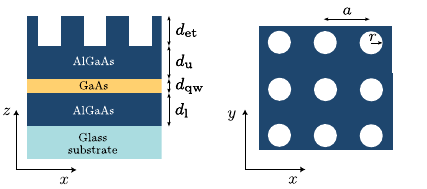}
\caption{Schematic representation of the simulated structure: a single QW layer is embedded in a partially etched photonic crystal lattice, the whole effective planar waveguide being deposited on a glass substrate. The vertical (i.e., along the $z$ direction) heterostructure parameters are defined in the left panel, while the main photonic crystal lattice parameters in the $(x,y)$ plane are given on the right panel.     
} \label{fig:structure_2D}
\end{figure}

\textit{Simulated structure parameters}---The calculations shown throughout this work refer to a specific realization of the generic patterned multilayer structure with 2D active materials represented in Fig.~1(a). In particular, we hereby consider a purely GaAs quantum well (QW) layer  embedded between AlGaAs barrier layers, and the whole structure is deposited on top of 
a glass substrate, which is reported in Fig.~\ref{fig:structure_2D} with a definition of all the characteristic parameters employed in the numerical simulations. The top part of the upper AlGaAs layer is partially patterned with a square lattice of circular air holes. The lattice period is  $a=\SI{250}{\nano\meter}$, and the hole radius is taken to be $\SI{80}{\nano\meter}$.
The layer thicknesses, defined in Fig.~\ref{fig:structure_2D}, are
$d_l=\SI{100}{\nano\meter}$,
$d_{\mathrm{QW}}=\SI{20}{\nano\meter}$,
$d_u=\SI{100}{\nano\meter}$, and
$d_{\mathrm{et}}=\SI{80}{\nano\meter}$.
For the GME simulations, the background dielectric constants are taken as
$\varepsilon_b(\mathrm{glass})=2.1$,
$\varepsilon_b(\mathrm{GaAs})=12.80$, and
$\varepsilon_b(\mathrm{AlGaAs})=11.02$.
The exciton is assumed with an effective mass  $M=0.2\,m_0$, and it is located at the midpoint of the QW layer. Its resonance energy is $E_{\mathrm{exc}}=\SI{1.584}{\electronvolt}$, and its oscillator strength per unit area is $f/S=\SI{e17}{\per\square\metre}$, as appropriate for GaAs excitons in a \SI{20}{\nano\meter} thick QW. More details on the input parameters for the RCWA benchmark simulations are given in Sec.~S4 of the SM.

\begin{figure*}[t]
\centering
\includegraphics[width=\textwidth]{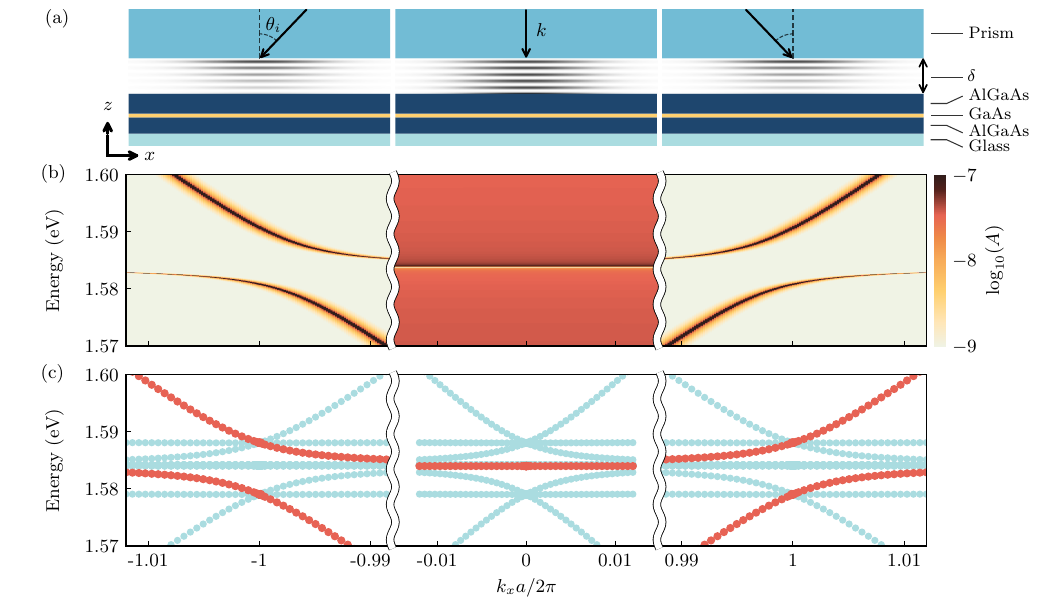}
\caption{{(a) Geometry used to excite radiative excitons and guided polaritons in a homogeneous dielectric slab. The structure is based on the one shown in Fig.~\ref{fig:structure_2D}, setting $d_u=\SI{125}{\nano\meter}$ and $d_{\mathrm{et}}=0$ such that the upper AlGaAs layer is unpatterned and the waveguide forms a homogeneous slab. Radiative excitons are excited under near-normal incidence, whereas guided polaritons outside the radiative light cone are accessed through attenuated total reflection (ATR) by using a high-index prism with dielectric constant $\varepsilon_{\mathrm{prism}}=13$, separated from the slab by an air gap of thickness $\delta=\SI{250}{\nano\meter}$. In the RCWA simulations, a small imaginary component of the dielectric constant of the GaAs layer is introduced ($\Im(\varepsilon)=10^{-6}$), giving rise to clear absorption response.}
(b) Absorption spectra calculated using RCWA close to the Brillouin zone edge. Around $|k_x|\approx2\pi/a$, the ATR configuration reveals the guided polariton dispersion, with finite radiative losses induced by coupling to the prism. Near normal incidence ($k_x\approx0$), only the radiative exciton branch is visible, instead.
(c) Polariton dispersion calculated with Hopfield matrix diagonalization for the same wavevector ranges shown in panel (b). Since GME directly computes the photonic eigenmodes of the structure, the prism is not required to access guided polaritons. Although the slab is translationally invariant in the plane, the plane-wave expansion formally folds the exciton and photon dispersions into the same Brillouin zone. The physical modes corresponding to $\mathbf{G}=(0,0)$ are highlighted in red, coinciding with the branches observed in panel (b). }
\label{fig:guided}
\end{figure*}

\textit{Guided Polaritons and Radiative Excitons in a Homogeneous Slab}---We hereby benchmark the generalized Hopfield formalism against the simpler case of a homogeneous dielectric slab. This comparison provides an intuitive connection with the well-established physics of planar waveguides, and it further validates the implementation of exciton radiative losses within the generalized Hopfield theory. In a homogeneous planar waveguide, the guided polariton modes lie entirely outside the light cone, and they are therefore inaccessible by direct free-space excitation. To compare the generalized Hopfield formalism with the RCWA reflectivity spectra, we therefore employ an attenuated total reflection (ATR) configuration, as previously implemented, e.g., in Refs.~\cite{Liscidini2011,Zanotti_CPC_2024}. In this geometry, light is injected through a high-index prism separated from the waveguide by an air gap. The large refractive index of the prism extends the accessible in-plane wavevector range to $k_x=n_{\mathrm{prism}}(\omega/c)\, \sin\theta_i$,
allowing wavevectors beyond the air light line to be reached through evanescent coupling and thereby enabling excitation of the guided polariton modes. 

Figure~\ref{fig:guided} shows excellent agreement between the dispersive features obtained from RCWA absorption spectra and the eigenmode dispersion calculated directly by diagonalizing the generalized Hopfield matrix. The RCWA calculation reproduces both the free-exciton resonance within the radiative light cone and the guided polariton branch (accessed through the ATR configuration). These branches coincide with the unfolded $\mathbf{G}=(0,0)$ bands obtained from the Hopfield eigenmode dispersion, confirming that the generalized Hopfield formalism correctly captures the dispersion properties of both excitons and guided polaritons.
{In Sec.~S7 of the SM we show systematic results as a function of the etching depth ($d_{\mathrm{et}}$), which interpolate between the homogeneous waveguide structure of Fig.~\ref{fig:guided} and the patterned one of Fig.~\ref{fig:structure_2D}. 
}

\onecolumngrid

\end{document}